\documentstyle[twoside,epsf,fleqn,espcrc2]{article}

\def\bc{\begin{center}}
\def\ec{\end{center}}
\def\be{\begin{equation}}
\def\ee{\end{equation}}
\def\myappendix{\par
 \setcounter{section}{0}
 \setcounter{subsection}{0}
 \setcounter{equation}{0}
 \setcounter{table}{0}
 \def\appendixname{Appendix}
 \def\appesection{\setcounter{equation}{0}\section}
 \def\@thesection{\Alph{section}}
 \def\thesection{\appendixname\hskip 1.10ex\Alph{section}}
 \def\thesubsection{\@thesection.\arabic{subsection}}
 \def\theequation{\@thesection.\arabic{equation}}
 \def\thetable{\@thesection.\arabic{table}}}

\newcommand{\beq}{\begin{equation}}
\newcommand{\eeq}{\end{equation}}
\newcommand{\beqn}{\begin{eqnarray}}
\newcommand{\eeqn}{\end{eqnarray}}

\def\vdir{v\kern-7.8pt\Big{/}}
\def\pdir{p\kern-7.8pt\Big{/}}

\title{Matrix elements of four-fermion operators in the HQET.}

\author{V.~Gim\'enez\address{Dep. de Fisica Teorica and IFIC, Univ. de Valencia,\\
Dr. Moliner 50, E-46100, Burjassot, Valencia, Spain.}}
\begin{document}
\begin{abstract}
The $B$-$\bar B$ mixing, $B$-meson lifetimes, the $B_{s}$-$\bar{B}_{s}$
lifetime difference and SUSY FCNC effects in $\Delta B=2$ processes are very
important measurable quantities in $B$-meson phenomenology whose theoretical
predictions depend on unknown matrix elements of several four--fermion
operators. We present preliminary results for the matrix elements of the 
relevant four--fermion operators computed on a sample of $600$ lattices
of size $24^3\times 40$ at $\beta=6.0$, using the SW--Clover action for light
quarks with rotated light quark propagators and the lattice version of 
the HQET for heavy quarks.
As a necessary ingredient of the calculation, we also present results for the
next-to-leading order matching of the full theory to the lattice HQET (one-
and two--loop anomalous dimensions, one--loop QCD-HQET matching coefficients
and one--loop continuum-lattice HQET matching coefficients).
\end{abstract}
\maketitle
\section{Introduction and Motivation.}
\label{motiv}
$B$--hadron decays are a very important source of information on the physics
of the standard model (SM) and {\it beyond}. In many important cases, {\it long
distance} strong contributions to these processes can be separated into matrix 
elements of local operators. Lattice QCD can then be used to compute these 
non--perturbative parameters from first principles. 
A list of some four--fermion operators relevant to $B$--meson phenomenology
is the following:\\
\begin{itemize}
\item the operator $O_{LL}=\bar b\, \gamma^{\mu}\, (1-\gamma_{5})\, q\, 
              \bar b\, \gamma_{\mu}\, (1-\gamma_{5})\, q$ ( $q$ denotes a light
              quark $u$, $d$ or $s$), which determines the
              theoretical prediction of the $B^0$-$\bar{B}^{0}$ mixing
              in the SM, together with 
$O^{S}_{LL(RR)}=\bar b\, (1\mp\gamma_{5})\, q\, \bar b\, (1\mp\gamma_{5})\, q$,
$O_{RR}$ and $O^{S}_{LR}$, with obvious notation, parametrize SUSY effects 
in $\Delta B=2$ transitions \cite{susy}. Furthermore, $O_{LL}$ and $O^{S}_{LL}$ determine
the $B_{s}$ width difference $\Delta \Gamma_{B_{s}}$ \cite{deltabs};
\item $Q_{LL}=\bar b\, \gamma^{\mu}\, (1-\gamma_{5})\, q\, 
              \bar q\, \gamma_{\mu}\, (1-\gamma_{5})\, b$, together with 
              $Q^{S}_{LR}$, parametrize
              spectator effects in the $\tau_{B}$ and $\tau_{B_{s}}$ lifetimes
              \cite{taubbs};
\item finally, the corresponding operators with a $t^{a}$ (the generator of the
SU(3) group) insertion: 
$Ot_{LL}$, $Ot^{S}_{LL}$ and $Ot^{S}_{LR}$ contribute to SUSY effects in 
$\Delta B=2$ transitions \cite{susy}, and $Qt_{LL}$ and $Qt^{S}_{LR}$ to spectator effects 
in the B-meson lifetimes \cite{taubbs}.                            
\end{itemize}

Our aim is to non-perturbatively evaluate the matrix elements 
between $B$--meson states 
of all the operators relevant to phenomenology using lattice simulations
of the HQET for the $b$--quark.
The procedure to perform the transition from QCD to lattice HQET, 
a combination of analytic and numerical calculations, consists in the four 
steps described in detail below.

\section{First step: the continuum QCD -- HQET matching.}
\label{firststep}

The starting point is to express the QCD operators as linear combinations of
the HQET ones in the continuum at a given scale, say, $\mu=m_{b}$.
To do so, some amplitudes of the relevant operators between 
appropriate external $B$-meson states are evaluated to one--loop order 
both in QCD and HQET. 
Once the amplitudes have been renormalized using the same scheme in both 
theories, the QCD amplitud is expanded in powers of
$1/m_{b}$ to lowest order. Finally, the coefficients of the matching can be
obtain by subtraction at the scale $\mu$.
It should be stressed that the matching is renormalization scheme
dependent. Notice also that new effective operators can be generated.
 
For $O_{LL}$, the matching was determined by Flynn {\it et al}
using NDR \cite{flynn} and by Gimenez \cite{mio2} using DRED. 
Now, we have extended the calculation to all the relevant operators in section 1. 
As we said before, new effective operators arise at one-loop
order; i.e.~the operator $O^{S}_{LL}$ is expressed as a linear combination of
two HQET operators $\hat{O}^{S}_{LL}$ and $\hat{O}_{LL}$.
We refer the reader to ref.\cite{prep1} for details.

\section{Second step: running down to $\mu=a^{-1}$ in the HQET.}
\label{secondstep}

The HQET operators obtained in step 1 at the high scale $\mu=m_{b}$ are evolved
down to a lower scale $\mu=a^{-1}$, appropriate for lattice simulations,
using the HQET renormalization group equations. 
The running is governed by the corresponding anomalous dimension matrices. 
At NLO, one has to calculate both the one- and two-loop anomalous 
dimension matrices of the relevant effective operators. The former are
renormalization scheme independent but the latter are not. The scheme
dependence of the two-loop anomalous dimension matrices cancels with
the scheme dependence of the matching in step 1.

The calculation of the one-loop anomalous dimension matrix is not difficult
\be
\label{eq:dim0}
\hat{\gamma}^{(0)}\, =\, \left[ \begin{array}{llll}
-8&0&0&0\\
4/3&-8/3&0&0\\
0&0&-7&6\\
0&0&3/2&-7
\end{array} \right]
\ee
in the basis $\{ \hat{O}_{LL}$, $\hat{O}^{S}_{LL}$, $\hat{O}_{LR}$, 
$\hat{O}^{S}_{RL}\}$.

However, only the two-loop anomalous dimension of the operator $\hat{O}_{LL}$
is known. It was determined by Gimenez \cite{mio2} (see also 
ref.\cite{nlo}) using NDR and DRED. Now we have 
extended the calculation to all effective operators in the basis. Technically
this is a very hard task because it involves the evaluation of 49 two-loop
Feynman diagrams in the HQET. Moreover, the operators mix among them, see
eq.(\ref{eq:dim0}), and with evanescent operators under renormalization. 
However, some useful relations between the matrix elements of the anomalous
dimension matrix can be obtained and used as a check of the calculation
using a renormalization prescription preserving Fierz symmetry.
We skip all details and refer the reader to ref.\cite{prep1}.

\section{Third step: continuum--lattice HQET matching.}
\label{thirdstep}

Having obtained the continuum HQET operators at the scale $\mu=a^{-1}$,
they are expressed as a linear combination of lattice 
HQET operators at this scale. The procedure is very similar to step 1: 
one matches two amplitudes at one-loop order, one 
in the continuum HQET and the other in the lattice HQET. 
However, some subtle points to remember are the following:
\begin{enumerate}
\item the matching has to be calculated using the same 
action used in the numerical simulation; the matching coefficients
depend on the action used in the simulation.
\item Due to the breaking of chiral symmetry induced by the Wilson term for light
quarks, the original operators can mix with new lattice operators.
\item The matching, as in step 1, depends on the renormalization procedure
used to define the operators in the continuum HQET, which cancels with 
the remaining scheme dependence of the running in step 2.
\end{enumerate}

The lattice counterpart of the operator $\hat{O}_{LL}$ is already known.
It was determined by Flynn {\it et al} \cite{flynn} for the Wilson action 
and by Borrelli and Pittori \cite{bp} for the SW-Clover action. Again, we have
extended the calculation to all the HQET operators in the basis of
eq.(\ref{eq:dim0}). Our results for $\hat{O}_{LL}$ agree with 
ref.\cite{flynn} but disagree with ref.\cite{bp} in the sign of 
the contribution proportional to $O^{latt}_{N}$ (see below) of the group of 
Feynman diagrams with radiative corrections between light quark legs only.
We refer the reader to ref.\cite{prep2} for details.
 
\begin{figure}[t]
\vspace{9pt}
\begin{center}\setlength{\unitlength}{1mm}
\begin{picture}(55,40)
\put(12,-10){\epsfbox{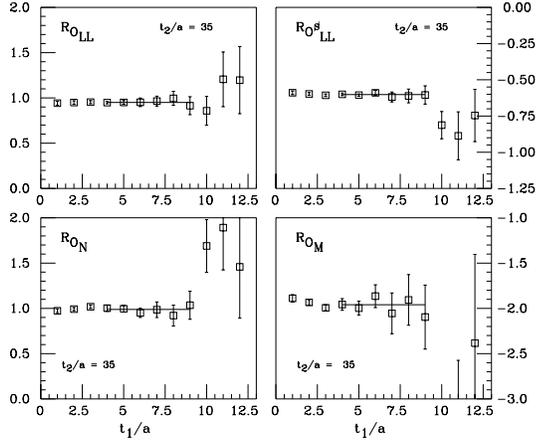}}
\end{picture}
\end{center}
\caption{\it{The ratios $R_{O_{LL}}$, $R_{O^{S}_{LL}}$, $R_{O_{N}}$ and 
$R_{O_{M}}$ computed at $K=0.1432$.}}
\label{fig:fourfig}
\end{figure}

As we said before, new lattice operators arise
\begin{eqnarray}
O^{latt}_{N} & = & O^{latt}_{LR}\, +\, O^{latt}_{RL}\, +\, 2\, (\, 
O^{latt\, S}_{LR}\, +\, O^{latt\, S}_{RL}\, )\nonumber\\
O^{latt}_{M} & = & 3/2\, (\, O^{latt}_{LL}\, +\, O^{latt}_{RR}\, )\nonumber\\
&+& 4\, (\, O^{latt\, S}_{LL}\, +\, O^{latt\, S}_{RR}\, )\\
O^{latt}_{P} & = & O^{latt}_{LR}\, +\, O^{latt}_{RL}\, +\, 6\, (\, 
O^{latt\, S}_{LR}\, +\, O^{latt\, S}_{RL}\, )\nonumber\\
O^{latt}_{Q} & = & O^{latt}_{LL}\, +\, O^{latt}_{RR}\, +\, 8\, (\, 
O^{latt\, S}_{LL}\, +\, O^{latt\, S}_{RR}\, )\nonumber
\end{eqnarray}
We found that $O^{latt}_{LL}$ mixes with $O^{latt}_{RR}$ and $O^{latt}_{N}$,
$O^{latt}_{LR}$ with $O^{latt\, S}_{RL}$ and $O^{latt}_{M}$,
$O^{latt\, S}_{LL}$ with $O^{latt}_{LL}$, $O^{latt}_{RR}$ and $O^{latt}_{P}$
and finally $O^{latt\, S}_{LR}$ mixes with 
$O^{latt}_{RL}$ and $O^{latt}_{Q}$.
We refer the reader to ref.\cite{prep2} for details.
 

\section{Fourth step: lattice computation of the matrix elements.}
\label{fourthstep}

In order to obtain the B--parameters of the relevant operators, we simulate on
the lattice the ratio (see, for example, ref.\cite{gm})
\be
R_{O_{i}}(t_{1},t_{2}) \, =\, 
\frac{C_{O_{i}}(-t_{1},t_{2})}{8/3\, C(-t_{1})\, C(t_{2})}
\ee
where $C_{O_{i}}(t_1, t_2)$ is the three-point correlation
function  with an insertion of the lattice operator $O_{i}$ and $C(t)$ is the
two-point correlation function for the $B$--meson. 
To extract the $B$-parameter, we search for a plateau in $t_{1}$ at 
fixed (and large) $t_{2}$. We observe,
see figs.1 and 2, good plateaux over large time--distances for all operators.
This makes us confident that the lightest meson state has been isolated. 
Furthermore, our results are almost independent of the light quark mass,
therefore they can be safely extrapolated to the chiral limit.
From figs.1 and 2,  we see that the correction given by the subleading 
operators may be important and cannot be neglected.
Using the results from steps 1 to 4, we can evaluate the
$B$--parameters of all operators relevant to phenomenology. For example, for
the renormalization scale independent $B$--parameter of $O_{LL}$ we found 
$\hat{B}_{B_{d}} =  1.14 \pm 0.16$ and 
$\hat{B}_{B_{s}}/\hat{B}_{B_{d}} = 1.01 \pm 0.01$, \cite{gm}. The calculation
of the other $B$--parameters is in progress and will be published elsewhere
\cite{prep2}. 

\begin{figure}[t]

\vspace{9pt}
\begin{center}\setlength{\unitlength}{1mm}
\begin{picture}(55,40)
\put(12,-10){\epsfbox{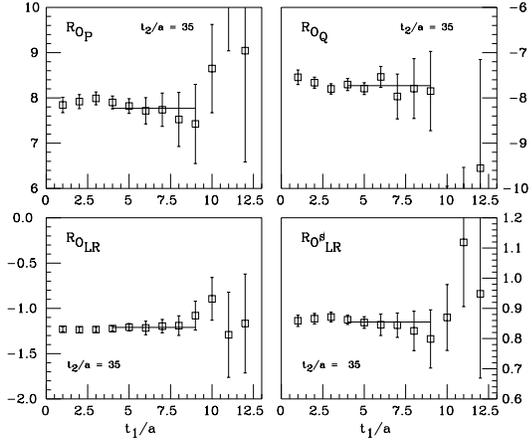}}
\end{picture}
\end{center}
\caption{\it{The ratios $R_{O_{P}}$, $R_{O_{Q}}$, $R_{O_{LR}}$ and
$R_{O^{S}_{LR}}$ computed at $K=0.1432$.}}
\label{fig:fourfig2}
\end{figure}

\end{document}